\documentclass[draft]{agujournal2019}
\usepackage{url} 
\usepackage{lineno}
\usepackage[inline]{trackchanges} 
\usepackage{soul}

\usepackage{natbib}
\usepackage{amsmath}
\usepackage{breqn}
\usepackage{pdfpages}
\newcommand{\track}[1]{\textcolor{black}{#1}}

\draftfalse

\journalname{JGR Planets}

\begin{document}

%
%


\title{Rapid hydrofracture of icy moon shells: insights from glaciology}

%
%




\authors{Robert Law\affil{1,2}}


\affiliation{1}{University of Bergen, Bergen, Norway}
\affiliation{2}{Bjerknes Centre for Climate Research, Bergen, Norway}




\correspondingauthor{Robert Law}{robert.law@uib.no}


\hfill \break
\hfill \break

\begin{keypoints}
\item \track{Hydrofracture in ice drives water-filled cracks downward due to the greater density of water} 
\item Rapid hydrofracture is possible for realistic icy-moon stress configurations given a supply of meltwater and an initial fracture
\item \track{Hydrofracture allows surface-interior material transfer with implications for habitability and may be associated with chaos terrain genesis}
\end{keypoints}

%
%

%
%


\begin{abstract}
Europa’s surface exhibits many regions of complex topography termed `chaos terrains’. One set of hypotheses for chaos terrain formation requires upward migration of liquid water from perched water bodies within the icy shell formed by convection and tidal heating. However, consideration of the behavior of terrestrial ice sheets suggests the upwards movement of water from \track{englacial} water bodies is uncommon. Instead, rapid downwards hydrofracture from supraglacial lakes — unbounded given a sufficient volume of water — can occur in relatively low tensile stress states given a sufficiently deep initial fracture due to the negative relative buoyancy of water. I suggest that downwards, not upwards, fracture may be more reasonable for perched water bodies but show that full hydrofracture is unlikely if the perched water body is located beneath a mechanically strong icy lid. However, full hydrofracture is possible in the event of lid \track{break up} over a perched water body and likely in the event of a meteor impact that generates sufficient meltwater and a tensile shock. This provides a possible mechanism for the transfer of biologically important nutrients to the subsurface ocean and the formation of chaos terrains. 
\end{abstract}

\section*{Plain Language Summary}
Jupiter's moon Europa has a subsurface ocean surrounded by an icy shell that features numerous `chaos terrains' of jumbled topography. Most theories for the formation of chaos terrains focus on the upwards motion of water pooled within the icy shell, but this contrasts with the rapid downwards motion of water in ice sheets on Earth. This paper considers the rapid downwards motion of water on Europa and other icy moons through water-driven fracture (hydrofracture), finding that it is a possibility if a large water body is not covered by a strong lid. This has implications for delivery of biologically important material from the moon's surface to its ocean, and may help explain the formation of chaos terrains.

%
%

%


%
%
%
%

\section{Introduction}
The presence of liquid water in icy moons beyond the asteroid belt was hypothesized by \citet{Lewis1971SatellitesNature} and supported by surface images and magnetic field readings from the Voyager, Galileo, \track{and} Cassini satellite missions \citep{Carr1998EvidenceEuropa, Khurana1998InducedCallisto, Porco2006CassiniEnceladus, Lorenz2008TitansWinds, Hendrix2019TheWorlds}. The presence of liquid water in these moons offers the tantalizing possibility of \track{exobiology}, which focussed research into whether the other basic requirements of chemical energy to drive metabolism and organic matter may also be met (e.g. \citealp{Kargel2000EuropasLife, Sephton2018HowMoons}), with Jupiter’s moon Europa receiving particular attention (e.g. \citealp{Chyba2001PossibleEuropa, NASA2017EuropaReport}).

One pathway for the possibility of life on Europa requires the exchange of oxidants and organics produced on the moon’s surface with the subsurface ocean (e.g. \citealp{Delitsky1998IceSatellites, Carlson1999HydrogenEuropa, Chyba2000EnergyEuropa, Chyba2001PossibleEuropa}), placing an emphasis on `chaos terrains’ as possible locations for \track{localized melting} of Europa’s icy shell \citep{Carr1998EvidenceEuropa, Thomson2001EvidenceFlux} \track{that may facilitate surface-ocean exchange}. However, direct contact between a thin layer of brittle surface ice and an underlying ocean requires an unrealistic amount of thermal energy to locally melt Europa’s icy shell \citep{Collins2000EvaluationEuropa, Goodman2004HydrothermalFormation}, the thickness of which is poorly constrained but may exceed 30 km \citep{Schenk2002ThicknessShapes, Sotin2002Europa:Melting, Billings2005TheRidges, Fagents2022Cryovolcanism}. \track{Theories to explain chaos formation} therefore frequently invoke the presence of a near surface water body emplaced through convective upwelling (e.g. \citealp{Sotin2002Europa:Melting, Figueredo2002GeologyChaos, Pappalardo2004TheDiapirism, Schmidt2011ActiveEuropa}) \track{upwards fracture of ice} \citep{Walker2021PropagationCracks}, \track{or} sills \citep{Michaut2014DomesSills} that then reach the surface through brittle \track{fracture} (e.g. \citealp{Crawford1988Gas-drivenEuropa, Collins2000EvaluationEuropa, Manga2007PressurizedEnceladus, Luzzi2021CalderaMars, Singer2023ThinRingGraben}). Perched \track{(i.e. disconnected from the subsurface ocean)} water bodies are required in these chaos terrain formation scenarios because the negative buoyancy of water with respect to ice means fractures propagating upwards from the subsurface ocean would extend at most 90\% of the way through the shell before reaching hydrostatic equilibrium \citep{Crawford1988Gas-drivenEuropa}.

However, consideration of water bodies \track{over ice sheets on Earth (supraglacial lakes)} strongly suggests that the negative relative buoyancy of water presents a major problem to their stability. \track{Antarctic ice shelves feature supraglacial lakes up to around 1 km in diameter, which are observed to drain rapidly through hydrofracture} \citep{Scambos2000ThePeninsula, Robel2019AHydrofracture, Dunmire2020ObservationsSheet, Warner2021RapidAntarctica, Banwell2024ObservedAntarctica}. \track{Such lakes also} pepper the margins of the Greenland Ice Sheet\track{,} with 28-45\% readily observed to drain rapidly through hydrofracture \citep{Das2008FractureDrainage., Doyle2013IceSheet, Fitzpatrick2014AGreenland, Cooley2017ObservationSheet, Williamson2018Dual-satelliteGreenland, Chudley2019SupraglacialGlacier}. Theoretical considerations, \track{and in the case of Greenland resultant ice-sheet speed up due to basal lubrication,} show that such fractures will reach the \track{ice-ocean or} ice-bed interface given a sufficient supply of water \citep{Weertman1973CanGlacier, VanderVeen2007FractureGlaciers, Krawczynski2009ConstraintsSheets}. In these settings an initial fracture is required for drainage through hydrofracture to occur. This may come from tensile shock from nearby hydrofracture events \citep{Christoffersen2018CascadingFracture}, a transient strain rate increase as a result of changes in the subglacial hydrological system (e.g. \citealp{Poinar2021ChallengesScale}), \track{ice-shelf flexure} (e.g. \citealp{Banwell2013BreakupLakes}), or reactivation of an existing weakness \citep{Chudley2019SupraglacialGlacier} though it is not clear that there is a single dominant statistically supported explanatory mechanism \citep{Williamson2018ControlsApproach}. Notably, supraglacial lakes on the Greenland Ice Sheet form in depressions in the ice surface which are generally dominated by compressive stresses \citep{Doyle2013IceSheet, Stevens2015GreenlandSlip, Chudley2019SupraglacialGlacier}. 

Slow density-driven downward transport of temperate ice (ice at the pressure-dependent melting point containing a small percentage of liquid water) through an icy moon shell has been considered as a mechanism that would \track{limit the longevity} of perched water bodies \citep{Karlstrom2014Near-surfaceColumbia, Carnahan2022Surface-To-OceanEuropa, Hesse2022DownwardPercolation, Kalousova2024EvolutionTitan}. \track{Linear elastic fracture mechanics has furthermore been used to investigate downwards dry fracturing on Europa, and upwards fracturing from the ice-ocean interface} \citep{Lee2005MechanicsShell, Rudolph2009FractureShells, Craft2016FracturingEuropa, Poinelli2019CrevasseEuropa}. \track{However,} the possible implications of rapid \track{downwards} hydrofracture \track{from perched water bodies have} not thus far been explored. In this paper I cover two perched water body settings: one formed through convective and tidal heating (Fig. 1a) and a second formed by a meteorite impact (Fig. 1d). Each of these water bodies may then be separated from the ice surface by a mechanically strong ice layer (Figs. 1b, e) or a weak melange of ice, water, and slush (Figs. 1c, f)\track{, such as that commonly found in Greenlandic fjords (e.g. \citealp{Todd2014AreGreenland})}. Calculating the stresses governing hydrofracture in a simplified two-dimensional plane indicates that full-thickness hydrofracture is possible when a mechanically strong lid is not present, \track{provided there is} a sufficient supply of water. I suggest hydrofracture may be more likely in the event of a meteorite strike that also provides an initial fracture site. Such a mechanism may \track{be associated with} chaos terrain formation if violent drainage prompts surface collapse, \track{providing} a clear pathway for the transport of organics and oxidants into the subsurface ocean. I focus on Europa, but the results are applicable to any ice shell setting.  

\begin{figure}
\noindent\includegraphics[width=0.95\textwidth]{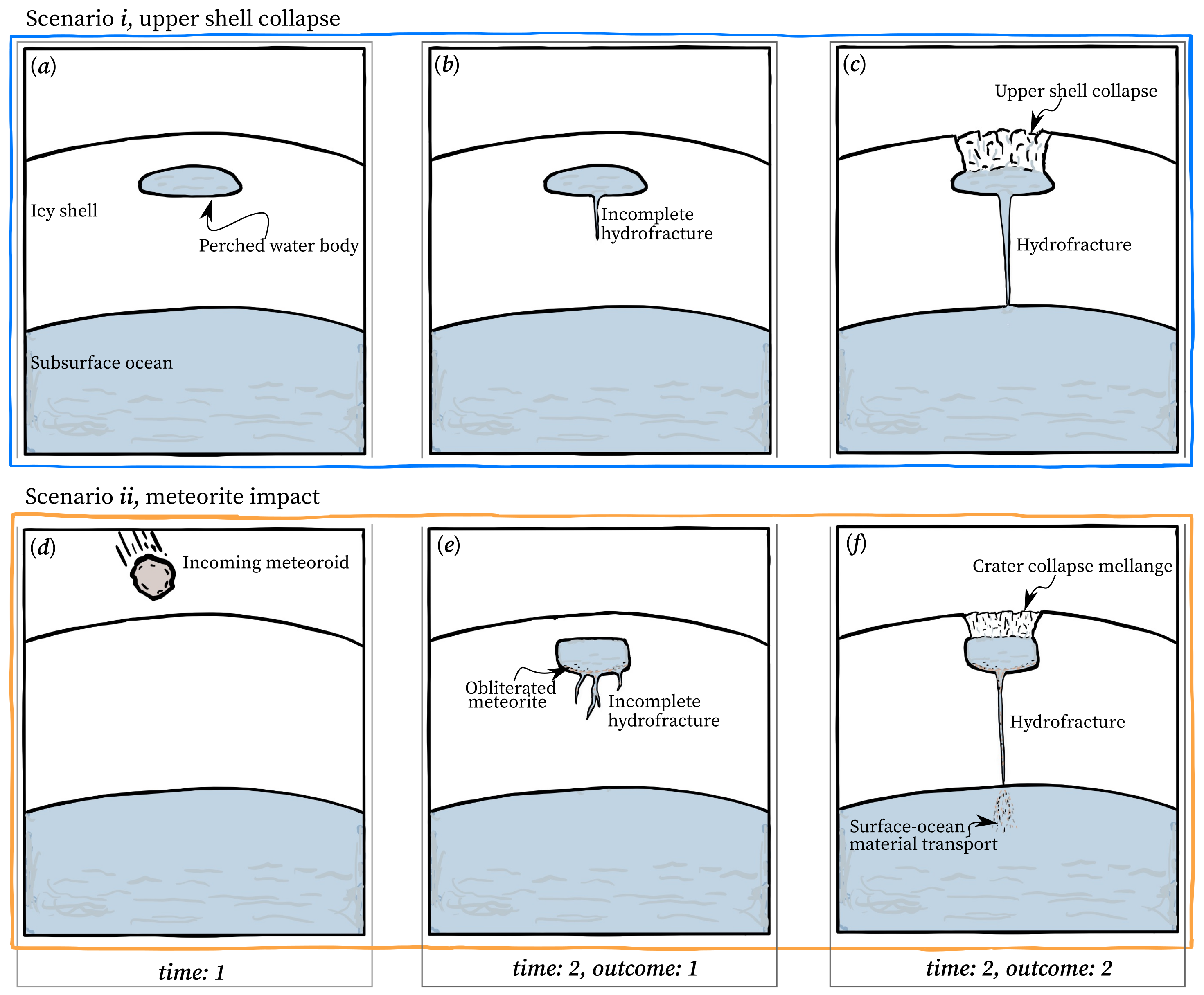}
\caption{Schematic of icy-shell hydrofracture scenarios covered in the text. (a) A perched water body as hypothesised to form through thermal convection or a sill. (b) Incomplete hydrofracture in a perched water body with a rigid lid. (c) Collapse of the upper shell and full hydrofracture. (d) Incoming meteoroid. (e) Incomplete hydrofracture following meteorite impact. (f) Full hydrofracture of meteorite-generated meltwater when the surface is not sealed following meteorite impact.}
\label{Fig:1}
\end{figure}

\subsection{Materials and methods}

Van der Veen (1998) calculates \track{water-filled} terrestrial crevasse propagation under linear elastic fracture mechanics which treats a material as containing small defects which affect its load-bearing capacity. If high stresses concentrate near a defect the defect may propagate and ultimately fracture \citep{Broek1982ElementaryMechanics}. In a purely elastic material, an unstable crack will propagate when the energy absorbed by a small expansion of the crack is exceeded by the energy released by a small increase in crack size \citep{Griffits1921VI.Solids}. This condition is met when the net stress intensity factor \track{at the fracture tip}, \(K_{(\textrm{net})}\) \track{(Pa m\textsuperscript{1/2})}, exceeds the fracture toughness of the material. Ice is not a purely elastic material (instead it has a non-linear viscoelastic rheology \track{and may be approximated as plastic at the fracture tip under very high stresses; \citealp{Walker2021PropagationCracks}}), but linear elastic fracture mechanics has proven effective in application to terrestrial ice sheets where temperatures are within 25 K of the pressure melting point throughout, and less than 10 K away from the pressure melting point in the zone of fracture initiation close to the surface (e.g. \citealp{vanderVeen1998FractureGlaciersb, Chudley2019SupraglacialGlacier, Law2021ThermodynamicsSensing}, covered further in the Discussion). \track{Nonetheless, full inclusion of viscous effects will still improve model realism \citep{Hageman2024IceLakes}}. The outer shells of icy moons are likely much colder (down to 100 K at the surface, but at the pressure melting point at the shell-ocean boundary) and therefore stiffer than terrestrial ice \citep{Goldsby2001SuperplasticObservations, Hussmann2002ThermalFlow, Ashkenazy2019TheEuropa} \track{with a correspondingly increased Maxwell time \citep{Lesage2022SimulationShells} making a focus on the elastic component only appropriate here, as in previous terrestrial and extraterrestrial studies (e.g. \citealp{Lee2005MechanicsShell, Poinelli2019CrevasseEuropa})}. \track{On Earth,} melt-formed conduits are also ubiquitous in temperate ice alongside fractures \citep{Fountain1998WaterGlaciers, Gulley2009AGlaciers} and surface water is still evacuated through several tens of meters of temperate ice at the bottom of an ice sheet \citep{Chudley2019SupraglacialGlacier, Law2021ThermodynamicsSensing}. 

The calculations below are separated into two parts. \textbf{First}, I use the essential equations from \citet{vanderVeen1998FractureGlaciersb} to calculate the stress states required for hydrofracture. \textbf{Second} I follow \citet{Weertman1973CanGlacier} and \citet{Krawczynski2009ConstraintsSheets} to calculate the volume of water required for hydrofracture. Considering first the stress states required for hydrofracture, I take a single water-filled fracture unaffected by surrounding fractures (Fig. 2) and consider only Mode I fracturing where the fracture opens symmetrically as a result of stresses normal to the fracture plane. \track{More closely spaced fractures will decrease the stress concentration at the fracture tip \citep{VanDerVeen2013FundamentalsDynamics}, but consideration of supraglacial lake drainages suggests that where an abundant water source is present fractures quickly coalesce with depth \citep{Chudley2021ControlsSheet}. In keeping with previous planetary and terrestrial studies (e.g. \citealp{Lee2005MechanicsShell, VanderVeen2007FractureGlaciers, Rudolph2009FractureShells, Craft2016FracturingEuropa}) and to conform with the assumptions of \citet{Broek1982ElementaryMechanics} we therefore exclude additional fractures to simplify calculations}. \track{Further, linear elastic fracture mechanics should only be considered strictly valid for cracks exceeding the area of plastic deformation at the fracture tip, which for this setting is around 2.5 m (Supporting Information; \citealp{Walker2021PropagationCracks}).}

The net stress intensity factor \(K^{(\textrm{net})}\) determines whether \track{a fracture propagates} if it exceeds the material fracture toughness, \track{and} is calculated as the sum of three processes\track{: the far-field tensile or compressive stress (as \(K^{(1)}\)); the hydrostatic stress (as \(K^{(2)}\)); and water pressure (as \(K^{(3)}\))}. \track{\(K^{(1)}\), \(K^{(2)}\), and \(K^{(3)}\)} can be superimposed without complication unlike if Mode II and Mode III openings were also considered \citep{vanderVeen1998FractureGlaciersb}, giving 

\begin{equation}
    K^{(\textrm{net})}=K^{1}+K^{2}+K^{3} \hspace{1mm} .
    \label{eq:net}
\end{equation}

The stress intensity factor corresponding to the tensile or compressive stress is given as 

\begin{equation}
    K^{(1)}=F(\lambda)R_{xx}\sqrt{\pi d}
\end{equation}

where \(\lambda =(d+h + a)/H, d\) (m) is the fracture depth, \(h\) is the height above the fracture surface (or water body surface if present) to the actual surface, \(a\) is the depth of the water body, and \(H\) is the ice shell thickness (Fig. 2). This differs slightly from the setup in \citet{VanderVeen2007FractureGlaciers} due to inclusion of a potentially deep perched water body and a potentially thick overlying ice lid \track{and in this context, \(a\) and \(h\) are taken to be small relative to \(H\)}. \(Rxx\) (Pa) is the resistive stress defined as 

\begin{equation}
    R_{xx}=\sigma_{xx}-L
\end{equation}

where \(\sigma_{xx}\) (Pa) is the full stress and \(L=-\rho_{i}g(H+h+ a-z)\) (m) is the lithostatic stress where \(\rho_i\) (kg m\textsuperscript{-3}) is the density of ice, \(g\) (m s\textsuperscript{-2}) is acceleration due to gravity and \(z\) is 0 at the base of the ice shell increasing upwards. Values of \(R_{xx}\)  \track{will} feature spatio-temporal variation \track{but are likely at least locally extensive at the ice-shell surface} (see Discussion for further details).  \(F(\lambda)\) \track{is the shape factor accounting for the fracture geometry, calculated as} 

\begin{equation}
    F(\lambda)=1.12 -0.23\lambda+10.55\lambda^2-21.72\lambda^3+30.39\lambda^4
\end{equation}

\track{following} \citet{vanderVeen1998FractureGlaciersb} \track{who draw from the compilation in} Broek (1982, \track{Table 3.1}). \track{This shape factor is intended for a fracture originating at the surface with a width to height ratio of 2 \citep{Broek1988TheMechanics}. Further consideration of a shape factor intended for a fracture originating in the middle of a plane, which reduces the value of \(F(\lambda)\) for all positive values of \(\lambda\), is Fig. S1.}
\nocite{Broek1982ElementaryMechanics}

\begin{figure}
\noindent\includegraphics[width=0.45\textwidth]{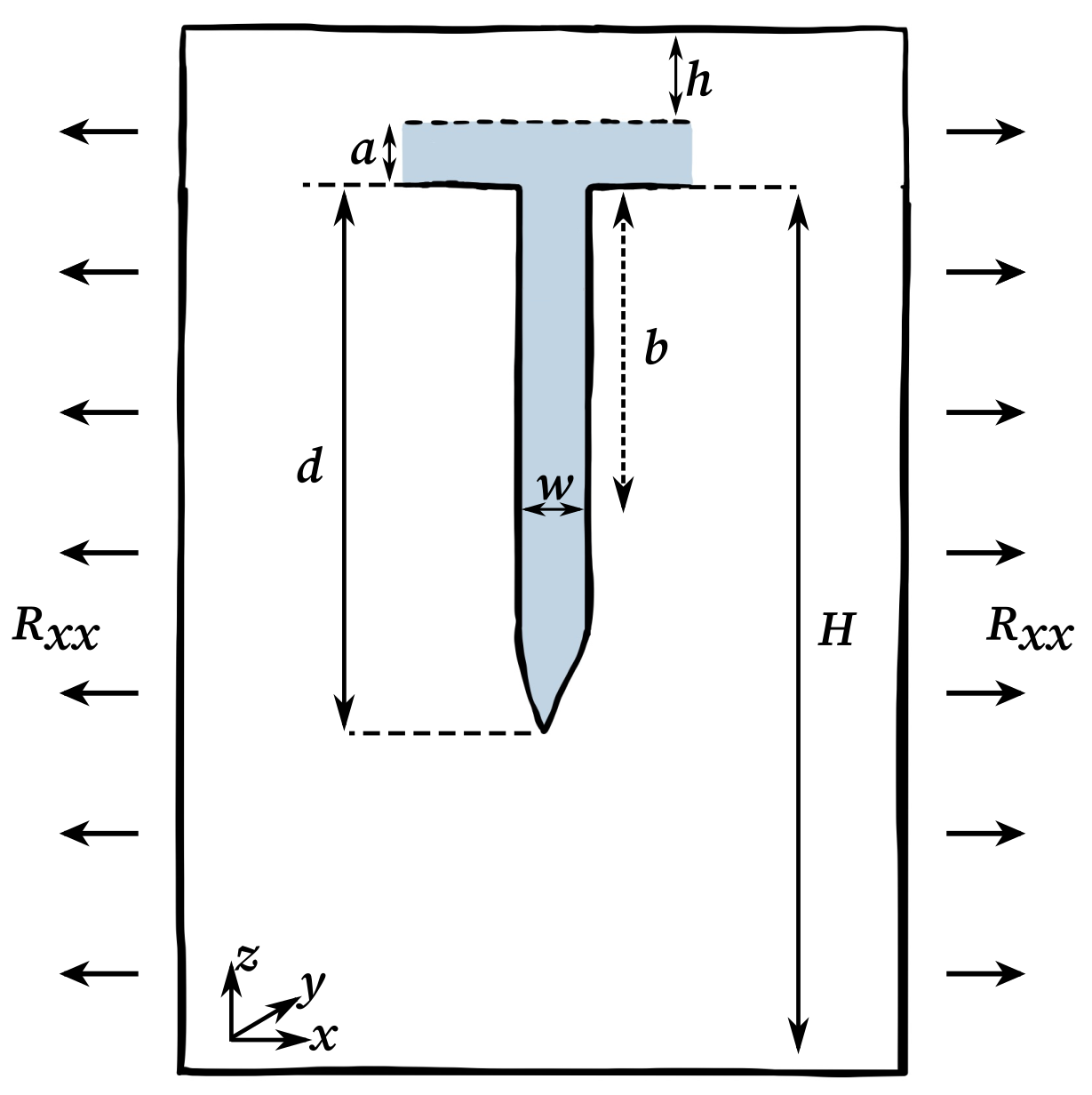}
\caption{Schematic of fracture calculations. See text for symbol definitions. The line at the top of the fracture is the fracture surface. The dashed line indicates the top of the perched water body.}
\label{Fig:2}
\end{figure}

\track{Next,} the net stress intensity factor from the overburden pressure is calculated by integrating the contributions from forces acting at different depths as

\begin{equation}
    K^{(2)}=-\frac{2g}{\sqrt{\pi d}}\int_{0}^{d}(\rho_i b+\rho_w a + \rho_i h)G(\gamma,\lambda)\textrm{d} b
    \label{eq:K2}
\end{equation}

where the \(\rho_w a\) and \(\rho_i h\) account for the weight of overlying water and overlying mechanically strong lid respectively and \(\rho_w\) is the density of water, \(\gamma =b/d\) and \(b\) is the depth from the fracture surface to the point in the fracture under consideration. I make the \track{simplifying} assumption that the ice lid is mechanically isolated from the water body and influences the hydrostatic stress (Eq. \ref{eq:K2}) but not the water pressure (Eq. \ref{eq:K3}). \track{In reality, downwards freezing of the ice lid may pressurize the water body beyond the assumptions used here, though any pressure will be released following initial fracture creation and ice displacement.} This simplification serves to illustrate the maximum fracture-limiting influence a lid overlying a perched water body may have. The \track{shape} function \(G(\gamma,\lambda)\) is

\begin{equation}
    G(\gamma,\lambda)=\frac{3.52(1-\gamma)}{(1-\lambda)^{3/2}}-\frac{4.35-5.28\gamma}{(1-\lambda)^{1/2}}+\left[ \frac{1.30-0.30\gamma^{3/2}}{(1-\gamma^2)^{1/2}} +0.83 -1.76\gamma \right]\times\left[1-(1-\gamma)\lambda \right]
\end{equation}

which \track{is} obtained by \citet{vanderVeen1998FractureGlaciersb} \track{from Tada et al. (2000, 2.29, p. 71)\footnote{Earlier editions of this textbook are difficult to locate so a more recent edition is cited. The equations are the same.}}. 

Last, the stress intensity factor from the water pressure within the fracture following \citet{vanderVeen1998FractureGlaciersb} is 

\begin{equation}
    K^{(3)}=\frac{2g}{\sqrt{\pi d}}\int_{0}^{d}(\rho_w b+\rho_w a)G(\gamma,\lambda)\textrm{d}b \hspace{1mm} .
    \label{eq:K3}
\end{equation}

\(K^{(2)}\) and \(K^{(3)}\) are then evaluated numerically to calculate \(K^{(\textrm{net})}\) at the given depth, \(d\), using Eq. \ref{eq:net}.

Separately to Eqs. 1-7, the volume of water required for a given crack propagation depth is calculated following \citet{Krawczynski2009ConstraintsSheets} and \citet{Weertman1973CanGlacier}. The fracture opening at a given depth is expressed as

\begin{dmath}
    w(b)=\frac{4\alpha\sigma}{\mu}\omega + 
    \frac{4\alpha\rho_i gd}{\pi \mu}\omega - 
    \frac{4\alpha\rho_w g}{\pi \mu}\omega\psi - 
    \frac{2\alpha\rho_i gb^2}{\pi \mu}ln \left( \frac{d+\omega}{d-\omega} \right) + 
    \frac{2\alpha\rho_w g(b^2-a^2)}{\pi\mu}ln \left|\frac{\psi+\omega}{\psi-\omega} \right| - \\
    \frac{4\alpha\rho_w gba}{\pi\mu}ln \left|\frac{a\omega+b\psi}{a\omega-b\psi} \right| + 
    \frac{4\alpha\rho_w ga^2}{\pi\mu}ln \left| \frac{\omega+\psi}{\omega-\psi} \right|
\end{dmath}

where \(\mu\) (Pa) is the shear modulus for ice, \(\alpha = 1-\nu\) where \(\nu\) is Poisson’s ratio, \(\omega=\sqrt{d^2-b^2}\) and \(\psi = \sqrt{d^2-a^2}\). In the application of \citet{Krawczynski2009ConstraintsSheets} and \citet{Weertman1973CanGlacier} the stress, \(\sigma\), at depth \(d\) is calculated as

\begin{equation}
    \sigma=R_{xx}-\frac{2\rho_i gd}{\pi} - \rho_w ga + \frac{\rho_w ga}{\pi}\textrm{sin}^{-1}\left(\frac{a}{d}\right)+\frac{2\rho_w g}{\pi}\psi \hspace{1mm} .
    \label{eq:krawczynski}
\end{equation}

This approach is a simplification of the stress calculations in Eqs. 1-7 from \citet{vanderVeen1998FractureGlaciersb}, but it is used here in keeping with \citet{Krawczynski2009ConstraintsSheets} given its proven efficacy in the setting in which it is applied. The fracture volume required for full thickness hydrofracture is then

\begin{equation}
    V_f=l\int_{0}^{H}w(b)\textrm{d}b
\end{equation}

which is evaluated numerically where \(l\) is the fracture length (along the \(y\) axis, into the page, and not displayed in Fig. 2). Although the water body is approximated as plane in Eqs. 1-7 to calculate the water body volume I consider a spherical perched water body of radius \(r\), the fracture length is then taken as \(l=\frac{4}{3}r\). On Earth the fracture may at first span the diameter of the water body \citep{Chudley2019SupraglacialGlacier}, but a more spherical geometry would result in a shorter fracture length. 
The radius of the required sphere is

\begin{equation}
    r=\sqrt{\frac{1}{\pi}\int_{0}^{H}w(b)\textrm{d}b} \hspace{1mm} .
\end{equation}

Values used in calculations are provided in Table 1.

\begin{table}
\caption{Values used in calculations. \(\nu\) from \citet{Krawczynski2009ConstraintsSheets}, \(\mu\) from \citet{Vaughan1995TidalMargins}.}
\centering
\begin{tabular}{l c l}
\hline
 Symbol  & Value & Meaning \\
\hline
  \(g\)  & 1.31 m s\textsuperscript{-2} & gravitational acceleration  \\
  \(\nu\)  & 0.3 & Poisson's ratio  \\ 
  \(\rho_i\)  & 917 kg m\textsuperscript{-3} & Ice density \\
  \(\rho_w\)  & 1000 kg m\textsuperscript{-3} & Water density  \\
  \(\mu\)  & 0.03, 0.3, 1.5 GPa & Shear modulus  \\
\hline
\multicolumn{2}{l}{}
\end{tabular}
\end{table}

\section{Results}

The calculations demonstrate that under certain circumstances hydrofracture from a perched water body to the subsurface ocean of Europa is possible. Fig. 3 shows the stress intensity experienced at the tip of the fracture against the fracture depth. If the stress intensity at the fracture tip exceeds the critical fracture toughness of ice the fracture will continue to propagate. If a mechanically strong lid is absent\track{, i.e. the lid is characterized as a weak layer of melange and slush,} the rate of change of \(K^{(3)}\) with depth exceeds that of \(K^{(2)}\) for all depths considered (Fig. S2) meaning that once the stress intensity factor at the fracture tip exceeds the critical fracture toughness the fracture will continue to propagate unbounded, limited only by the water supplying the fracture. An initial fracture depth of only 4.1 m under a tensile stress of 0.1 MPa is required for unstable hydrofracture if the water body is located on the surface. If an ice lid greater than around 10\track{0} m (dependent slightly on the value of \track{\(a\)}) is present that contributes to lithostatic but not hydrostatic pressure then the rate of change of \(K^{(2)}\) with depth exceeds that of \(K^{(3)}\) meaning fracture propagation is not favorable. Unreasonably deep initial fractures are required in the case of a surface water body and 0 MPa tensile stress but, once a fracture is established neutral far-field stress at depth would not lead to closure (Fig. S3).

\begin{figure}
\noindent\includegraphics[width=0.8\textwidth]{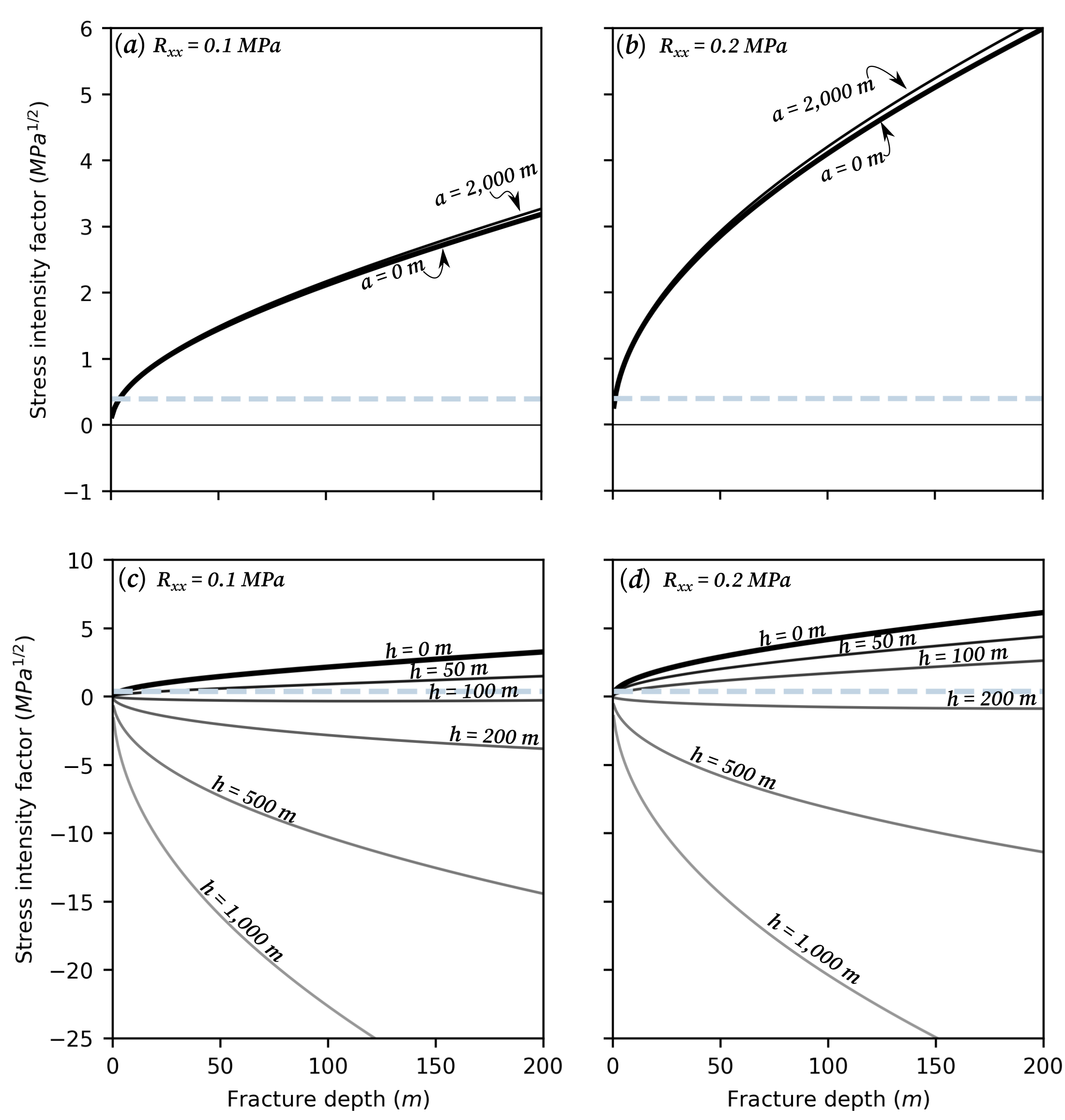}
\caption{Stress intensity factor as function of fracture depth for varying ice lid thickness, water body depth, and tensile stress. The blue dashed line is a reasonable upper limit for the fracture toughness of ice (0.4 MPa m\textsuperscript{1/2}) from \citet{vanderVeen1998FractureGlaciersb}. (a) and (b) the thickness of the overlying lid \(h\) = 0 m. (c) and (d) the depth of the water body \(a\) = 500 m. (a) and (c) tensile stress is 0.1 MPa. (b) and (d) tensile stress is 0.2 MPa. \(H\) = 30 km for all panels. Where the solid lines in panels (a-d) consistently increase above the fracture toughness unbounded hydrofracture can occur.}
\label{Fig:3}
\end{figure}

The volume of water required to facilitate a 30 km fracture under a shear modulus of 1.5 GPa is 38.8 km\textsuperscript{3} \track{, or 570 km\textsuperscript{3} for a shear modulus of 0.3 GPa} (Fig. 4). This is significantly below the 20,000-60,000 km\textsuperscript{3} melt lens estimate of \citet{Schmidt2011ActiveEuropa} and \track{around the same magnitude} of melt volumes modeled to form by \citet{Kalousova2024EvolutionTitan} in the event of moderately sized meteorite impacts. \track{Lower shear modulus values markedly increase the required water volume. Laboratory experiments suggest a shear modulus of \(\sim\)3.3 GPa for pure ice \citep{Gammon1983ElasticSpectroscopy}, which may decrease to \(\sim\)0.3-1.7 GPa for metamorphosed meteoric ice \citep{Vaughan1995TidalMargins, Godio2015TheMethods}. \citet{Nimmo2006NormalProperties} suggest 0.03-0.3 GPa for Europa based on analysis of normal faults, highlighting near-surface fracturing as a potential cause. A value at the upper end of Europa estimates may be more appropriate here however, as the fracture volume is mostly dependent on ice properties at depth within the shell. On Earth it is furthermore unclear the degree to which a fracture must maintain its width throughout propagation; by the time it is possible to conduct an investigation the passage is always narrow \citep{Catania2008CharacterizingSheet, Chudley2019SupraglacialGlacier}.} 

\begin{figure}
\noindent\includegraphics[width=0.45\textwidth]{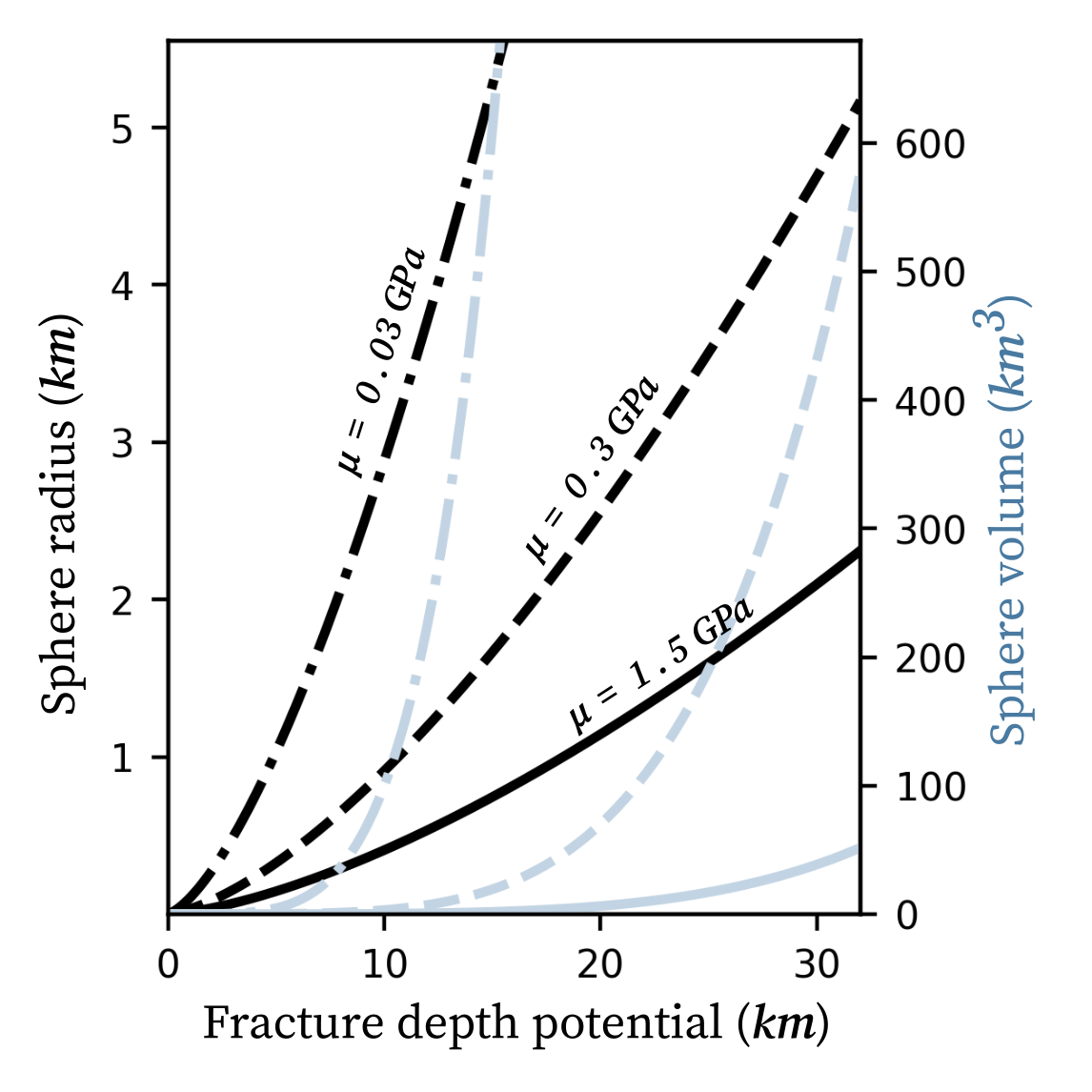}
\caption{Water volume required for a given fracture depth. Dashed line is for a shear modulus of 0.5 GPa and solid line for 1.5 GPa. Note that the upper bound shear modulus of 3.9 GPa, as used in \citet{Krawczynski2009ConstraintsSheets}}, is not included here. Black indicates required sphere radius, blue indicates sphere volume.
\label{Fig:4}
\end{figure}

\citet{VanderVeen2007FractureGlaciers} shows that the key control on the downward propagation of the fracture is the filling rate and that penetration velocity equals the rate of fracture filling to within a few percent. This is because \(K^{(2)}\) and \(K^{(3)}\) become the dominant terms in Eq. 7 giving to a good approximation

\begin{equation}
    d=\left( \frac{\rho_w}{\rho_i}\right)^{\frac{2}{3}}Qt
\end{equation}

where \(Q\) is the vertical fill rate (m s\textsuperscript{-1}) and \(t\) (s) is the time period in consideration. Hydrofracture could therefore be a very rapid phenomenon given a suitably placed large perched water body. Supraglacial lakes in Greenland take on the order of 2-3 hours to drain, but the water bodies have an aspect ratio of 1:100 making \(Q\) limited by lateral transport from the outer reaches of the lake \citep{Das2008FractureDrainage., Chudley2019SupraglacialGlacier}. As a first approximation, even if drainage took 24 hours through ice at 100 K, only 0.34 m of side wall freezing would occur (Text S1), negligible in the context of a surface opening of \(\sim\)200 m following Eq. 8.

\section{Discussion}

Figs. 3 and 4 demonstrate that full hydrofracture of Europa is possible given moderate tensile stresses, sufficient water to propagate the fracture, and \track{a lid comprised of an ice melange lacking functional mechanical strength}. Fractures are common across Europa’s surface (e.g. \citealp{Dombard2013FlankingEuropa}) but abundant surface water akin to the Greenland Ice Sheet is practically impossible given the very low temperature and pressure of the ice surface \citep{Ashkenazy2019TheEuropa}. Nonetheless, in this simple if unrealistic situation, hydrofracture can occur at tensile stresses much lower than the \(>\)20 MPa proposed by \citet{Crawford1988Gas-drivenEuropa} and \citet{Rudolph2009FractureShells} for full `dry’ shell fracture. \track{As for any planetary body, it is challenging to quantify the surface and depth-averaged stress state of Europa. However, estimates for the surface are around the 0.1-0.2 MPa, with regions of compression and extension both expected \citep{Pappalardo1998GeologicalShell, Hoppa1999FormationEuropa, Greenberg2002Tidal-tectonicCrust, Kattenhorn2006Fault-inducedMechanics}. Cooling and freezing of liquid water to the ice shell base may furthermore lead to extensive extension throughout the ice shell \citep{Nimmo2004StressesEuropa}. As with Antarctic ice shelves and the Greenland Ice Sheet, hydrofracture of perched water bodies may therefore only occur under spatially or temporally limited stress conditions.} 

However if, as widely hypothesized \citep{Fagents2022Cryovolcanism}, a water body is entombed then the additional lithostatic stress \track{and fracture stress configuration} makes rapid hydrofracture (or any \track{Mode I} fracture at all) unlikely. Though it is worth noting that mechanisms hypothesized to direct water upwards such as the pressurization through freezing of perched water bodies \citep{Manga2007PressurizedEnceladus, Steinbrugge2020BrineEuropa, Lesage2022SimulationShells, Quick2023ProspectsPlanets}, ought also to work downwards and will be aided in that case by density-difference driven fracture. But, if the lid is \track{comprised of weak melange} then its weight will contribute to the hydrostatic pressure and not the lithostatic pressure and hydrofracture again becomes possible. Two such scenarios may occur. 

First, \citet{Schmidt2011ActiveEuropa}, \citet{Walker2015IceEuropa}, and \citet{Luzzi2021CalderaMars} among others propose collapse of the upper shell above perched water bodies as a formation mechanism for chaos terrains. In several of these scenarios the blocks of ice are buoyantly supported by brine, water, and slush (Fig. 1c). This effectively sets \(h\) at zero, while increasing \(a\), meaning hydrofracture becomes a realistic possibility. It is unclear how long this lid state may persist before refreezing occurs \citep{Schmidt2011ActiveEuropa, Fagents2022Cryovolcanism}, but while it does only a small and temporary tensile stress and a very small initial fracture would be required to initiate rapid hydrofracture.  

Second, literature on the impacts of meteor impacts on icy moons is limited, but it appears reasonable that \track{an impactor} may provide three ingredients for hydrofracture — an initial fracture site, a tensile shock, and an abundant supply of impact-generated melt water \citep{Kalousova2024EvolutionTitan} while leaving an at least transiently mechanically weak area above the water body (Fig. 1f). A meteorite may also impact sufficiently close to an existing perched water body to trigger its hydrofracture\track{, though meteorite impacts have also been invoked as a potential cause of cryovolcanism \citep{Steinbrugge2020BrineEuropa}}. 
If density driven sinking of melt \citep{Carnahan2022Surface-To-OceanEuropa, Kalousova2024EvolutionTitan} actually prevents the formation of a perched water body, then a meteorite impact becomes the only remaining possible source \track{of a large water body}.

If a previously entombed water body drains, the drainage event will remove hydrostatic support and lead to extensive collapse of the overlying terrain. If the fracture remains open the water level would then be expected to stabilise at \(\sim\)90\% of the ice shell thickness \citep{Michaut2014DomesSills}, though viscous deformation and freezing may be expected to close the fracture over a short period (years or less) after active drainage has ceased \citep{Andrews2022ControlsModel}. It is worth noting that in Greenland the large ice blocks generated in initial hydrofracture, with more expected if an icy lid is present \citep{Russell1993SupraglacialGreenland}, may become wedged at the fracture surface (Fig. S4) so while water may leave the site quickly, solid ice blocks should remain. This supports a more neocatastrophist view of chaos terrain formation than gradual collapse of the upper shell \citep{Luzzi2021CalderaMars}.

Last, it is challenging to be definitive about the thermal state of Europa’s icy shell, but several studies suggest a warm and uniform convecting interior, within around 10 K of the pressure melting point (e.g. \citealp{McKinnon1999ConvectiveShell, Hussmann2002ThermalFlow, Showman2004NumericalFeatures, Mitri2005ConvectiveconductiveEuropa}). These temperatures are warm relative to an average surface temperature of around 100 K \citep{Ashkenazy2019TheEuropa}, but their correspondingly lower relative viscosity should not present a problem to elastic fracture given full-thickness hydrofracture through ice of this temperature is common in the Greenland Ice Sheet (e.g. \citealp{Doyle2018PhysicalGreenland, Law2021ThermodynamicsSensing}). More intriguing is the possible influence of temperate ice on water transport within Europa’s shell. \citet{Carnahan2022Surface-To-OceanEuropa} and \citet{Kalousova2024EvolutionTitan} do not account for the formation of conduits within the icy shell, yet there are no temperate terrestrial glaciers where there is not some degree of englacial drainage system formation (e.g. \citealp{Hooke1989EnglacialReview}). A body of temperate ice within an icy shell would lack a supply of surface meltwater, but tidal or convective heating may still produce water internally (e.g. \citealp{Gaidos2000TectonicsEuropa}). Given that the melt chambers hypothesized by \citet{Carnahan2022Surface-To-OceanEuropa} reach \(>\)5 km in height and feature melt fractions \(>\)5\%, it seems reasonable that transport through a drainage system may lead to pooling of water at the transition between temperate and cold ice if fracture does not occur. Further integrating understanding of englacial hydrology with icy moon mechanics can advance understanding of these mechanisms. 

\section{Conclusions}

This paper demonstrates that in the presence of a mechanically weak lid only relatively small tensile stresses, and achievable water volumes, are required for full hydrofracture of Europa\track{'s ice shell}. This has direct application to other icy moons with different gravities and stress regimes, which may also have conditions facilitating full-thickness hydrofracture. It also provides a clear mechanism for the delivery to the subsurface ocean of biologically important materials created on the surface of Europa and delivered by incoming meteorites and may provide an explanation for the formation of chaos terrains. Future research directions include numerical modeling of the spatio-temporal variation of stress configurations around varied perched water body geometries \citep{Craft2023ImpactWorlds, Hageman2024IceLakes} and as a result of meteorite impacts, and improving understanding of the geomorphological implications of rapid hydrofracture. Hopefully, this paper opens further interdisciplinary pathways \citep{Garcia-Lopez2017GlaciersWorlds, Rossi2023DeformationAnalogues} between glaciology and the study of icy moons.

\section*{Open Research Section}
\track{The} data supporting the conclusions is derived from the calculations outlined in the text. \track{Plotting and analysis scripts are available at \cite{Law2024ReplicationV1}.}

\acknowledgments
Thanks to Craig Walton and Claire Guimond for introducing me to this field and Andreas Born for letting me take a bit of time away from the day job. \track{Thanks to two anonymous reviewers for very helpful comments.} I acknowledge funding from \track{The Research Council of Norway} (Norges Forskningsråd) grant 314614. There are no conflicts of interest with respect to the results of this manuscript.
%
\bibliography{references} 
%


%
%
%
%
%

\end{document}